\documentstyle[12pt]{article}
\author{V. A. Tsokur, Yu. M. Zinoviev \thanks{E-mail address:
        ZINOVIEV@MX.IHEP.SU}\\
        {\it Institute for High Energy Physics } \\
        {\it Protvino, Moscow Region, 142284, Russia }}
\title{Spontaneous supersymmetry breaking \\
       in $N=4$ supergravity with matter}
\date{}
\begin{document}
\thispagestyle{empty}
\maketitle
\begin{abstract}
  In this paper we consider the problem of spontaneous supersymmetry
breaking in $N=4$ supergravity interacting with vector multiplets. We
start with the ordinary version of such model with the scalar field
geometry $SU(1,1)/U(1)\otimes SO(6,m)/SO(6)\otimes SO(m)$. Then we
construct a dual version of this theory with the same scalar field
geometry, which corresponds to the interaction of arbitrary number of
vector multiplets with the hidden sector, admitting spontaneous
supersymmetry breaking without a cosmological term. We show that
supersymmetry breaking is still possible in the presence of matter
fields.
\end{abstract}

\newpage
\setcounter{page}{1}

\section{Introduction}

   In this paper we continue our investigation of the problem of
spontaneous supersymmetry breaking in extended supergravities started
in \cite{Tso94}, where $N=3$ case have been considered. For the $N=4$
supergravity the only natural candidate for the model of supergravity
--- vector multiplets interaction is the one with the scalar field
geometry $SU(1,1)/U(1)\otimes SO(6,m)/SO(6)\otimes SO(m)$ \cite{Der84}.
Such a model
has indeed been constructed some time ago
\cite{Ber85,Roo85,Roo85a,Roo85b}, in this the whole group $SO(6,m)$
turned out to be the symmetry of the Lagrangian and not that of the
equations of motion only. So in this aspect the $N=4$ supergravity
seems to be drastically different from $N=3$ theory, but it is
important to note that the $N=4$ supergravity multiplet itself
contains couple of scalar fields, describing the non-linear $\sigma
$-model
$SU(1,1)/U(1)$. This leads to the existence of the dual versions for
this theory, the most known cases being $SO(4)$ \cite{Das77} and
$SU(4)$ \cite{Cre77,Cre78} theories. But in turn, this should lead to
the existence of dual versions for $N=4$ supergravity --- vector
multiplet interaction as well. In \cite{Zin87} the hidden sector for
$N=4$ supergravity was constructed, which was the dual version for the
system $N=4$ supergravity with six vector multiplets and admitted the
spontaneous supersymmetry breaking without a cosmological term. In
this paper we start with the ordinary version for vector multiplets
interaction and then, using the fact that all dual versions have the
same scalar field geometry, we manage to construct the model, which
turns out to be the generalization of our hidden sector to the case of
arbitrary number of vector multiplets. Moreover, we show that the
spontaneous supersymmetry breaking is still possible in the presence
of matter fields.

\section{Ordinary version}

   Let us start with a well known model of vector multiplets --- $N=4$
supergravity interaction when the scalar fields describe the
non-linear $\sigma $-model $[SU(1,1)/U(1)]\otimes [SO(6,m)/SO(6)\otimes
SO(m)]$. We need
the following set of fields: graviton $e_{\mu r}$, gravitini $\Psi _{\mu
i}$,
$i=1,2,3,4$, vector fields $A_\mu {}^A$, $A=1,2,...m+6$,
$g_{AB}=diag(------,+...+)$, majorana spinors $\Omega _i{}^A$ and $\lambda
_\alpha {}^i$,
$\alpha =1,2$, real scalars $\Phi _a{}^A$, $a=1,2,...6$ and complex
scalars
$z^\alpha $. In this, the spinor and scalar fields must satisfy the
following constraints:
\begin{eqnarray}
 SU(1,1)/U(1) &:& \qquad z^\alpha  \bar{z}_\alpha  = -2, \qquad z^\alpha
\lambda _\alpha  = 0
\label{c1} \\
 SO(6,m)/SO(6)\otimes SO(m) &:& \qquad \Phi _a{}^A \Phi _{Ab} = - \delta
_{ab}, \qquad
 \Phi _a{}^A \Omega _A=0 \label{c2}
\end{eqnarray}

   In such formulation the theory have local $O(6)\otimes U(1)$
invariance,
where the corresponding covariant derivatives look like, e.g.:
\begin{eqnarray}
 D_\mu  \Phi _a{}^A &=& \partial _\mu  \Phi _a{}^A - (\Phi
_a\stackrel{\leftrightarrow } \partial _\mu  \Phi _b) \Phi _b{}^A,
\qquad   \Phi _a D_\mu  \Phi _b = 0 \nonumber \\
 D_\mu  z^\alpha  &=& \partial _\mu  z^\alpha  + \frac{1}{2} (\bar{z}
\partial _\mu  z) z^\alpha   \qquad
\bar{z}_\alpha  D_\mu  z^\alpha  = 0 \nonumber \\
 D_\mu  \Omega ^A &=& D_\mu ^g \Omega ^A + \frac{1}{4} (\Phi
_a\stackrel{\leftrightarrow } \partial _\mu  \Phi _b) \Sigma ^{ab}
\Omega ^A + \frac{1}{4} (\bar{z} \partial _\mu  z) \Omega ^A \label{d1}
\\
 D_\mu  \lambda _\alpha  &=& D_\mu ^g \lambda _\alpha  - \frac{1}{4} (\Phi
_a \stackrel{\leftrightarrow } \partial _\mu  \Phi _b) \Sigma ^{ab}
\lambda _\alpha  - \frac{1}{4} (\bar{z} \partial _\mu  z) \lambda _\alpha
\nonumber \\
 D_\mu  \eta  &=& D_\mu ^g \eta  + \frac{1}{4} (\Phi
_a\stackrel{\leftrightarrow } \partial _\mu  \Phi _b) \Sigma ^{ab} \eta  -
\frac{1}{4} (\bar{z} \partial _\mu  z) \eta  \nonumber
\end{eqnarray}

  For the case of abelian vector multiplets the full Lagrangian
(omitting four fermionic terms) has the form:
\begin{eqnarray}
 L_1 &=& - \frac{1}{2} R + \frac{i}{2} \varepsilon ^{\mu \nu \rho \sigma }
\bar{\Psi }_\mu  \gamma _5 \gamma _\nu  D_\rho
\Psi _\sigma  + \frac{i}{2} \bar{\Omega } \hat{D} \Omega  + \frac{i}{2}
\bar{\lambda } \hat{D} \lambda  +
\frac{1}{2} D_\mu  \Phi _a D_\mu  \Phi _a + \nonumber \\
 && + \frac{1}{2} D_\mu  z^\alpha  D_\mu  \bar{z}_\alpha  - \frac{1}{2}
\bar{\Omega } \gamma ^\mu  \gamma ^\nu
D_\nu  \Phi _a \bar{\tau }^a \Psi _\mu  - \frac{1}{2} \bar{\lambda
}_\alpha  \gamma ^\mu  \gamma ^\nu  D_\nu  z^\alpha  \Psi _\mu
\nonumber \\
 && - \frac{1}{4} \frac{1}{|K\bar{z}|^2} \left\{ (A_{\mu \nu })^2 + 2(\Phi
_a
A_{\mu \nu })^{2} \right\} - \frac{\gamma _5}{4} \left\{
\frac{\bar{M}z}{\bar{K}z} - \frac{M\bar{z}}{K\bar{z}} \right\}
(A_{\mu \nu } \tilde{A}_{\mu \nu }) + \nonumber \\
&& + \frac{1}{4} \bar{\Psi }_\mu  \frac{1}{K\bar{z}} (A^{\mu \nu } -
\gamma _5
\tilde{A}_{\mu \nu }) \Phi _a \bar{\tau }^a \Psi _\nu  + \frac{i}{4}
\bar{\Omega } \gamma ^\mu  (\sigma A)
\frac{1}{K\bar{z}} \Psi _\mu  - \nonumber \\
&& - \frac{i}{8} \bar{\lambda }_\alpha  \gamma ^\mu  \frac{\varepsilon
_{\alpha \beta } z^{\beta }}{\bar{K}z} (\sigma A) \Phi _a
\bar{\tau }^a \Psi _\mu  + \frac{1}{8}\bar {\Omega }^A (\sigma A) \Phi _a
\bar{\tau }^a
\frac{1}{\bar{K}z} \Omega ^A + \nonumber \\
 && + \frac{1}{4} \bar{\Omega } (\sigma A) \frac{\bar{z}_\alpha
\varepsilon ^{\alpha \beta }}{K\bar{z}} \lambda _\beta
\label{l1}
\end{eqnarray}
where
\begin{equation}
 \bar{K}z = (\bar{K}_\alpha  z^\alpha ),  \qquad  \bar{M}z =
(\bar{M}_\alpha  z^\alpha )
\end{equation}
and $K_\alpha $, $M_\alpha $ -- constant vectors, satisfying the
relations:
\begin{equation}
 K_\alpha  = \varepsilon _{\alpha \beta } \bar{K}^\beta , \qquad
(\bar{K}_\alpha  M_\alpha ) = \frac{1}{2}
\end{equation}

Here we introduced six antisymmetric matrices $(\tau ^a)_{ij}$ such that:
\begin{eqnarray}
 && (\bar{\tau }^a)^{ij} = \left[ (\tau ^a)_{ij} \right]^*= \frac{1}{2}
\varepsilon ^{ijkl} (\tau ^a)_{kl} \nonumber \\
 && (\tau ^a)_{ij} (\bar{\tau }^b)^{jk} + (\tau ^b)_{ij} (\bar{\tau
}^a)^{jk} = -
2 \delta _i{}^k \delta ^{ab}
\end{eqnarray}
Besides, we will need fifteen matrices
\begin{eqnarray}
 \left\{ \Sigma ^{[ab]} \right\}_i{}^j &=& \frac{1}{2} \left[ (\tau
^a)_{ik}
(\bar{\tau }^b)^{kj} - (\tau ^b)_{ik} (\bar{\tau }^a)^{kj} \right]
\end{eqnarray}
which are generators of the group $O(6)\approx SU(4)$.
Let us recall, that we use the $\gamma $-matrix representation where
majorana spinors are real, so that in all expressions with spinors
(including matrices $\tau ^a$ and $\Sigma ^{ab}$) matrix $\gamma _5$ plays
the role of
imaginary unit $i$. This leads, e.g.:
\begin{equation}
 \gamma _\mu  (\tau ^a)_{ij} = (\bar{\tau }^a)^{ij} \gamma _\mu , \qquad
(\Sigma ^{ab})_i{}^j \gamma _\mu  =
- \gamma _\mu  (\Sigma ^{ab})_j{}^i
\end{equation}

    The Lagrangian (\ref{l1}) is invariant under the following local
supertransformations:
\begin{eqnarray}
 \delta e_{\mu r} &=& i(\bar{\Psi }_\mu  \gamma _r \eta ) \nonumber \\
 \delta \Psi _\mu  &=& 2D_\mu  \eta  + \frac{i}{4\bar{K}z} (\sigma A) \Phi
_a \tau ^a \gamma _\mu  \eta  \nonumber \\
 \delta A_\mu {}^A &=& (\bar{\Psi }_\mu  \bar{K}z \Phi _a{}^A \bar{\tau
}^a \eta ) - i(\bar{\lambda }_\alpha
\gamma _\mu  \Phi _a \bar{\tau }^a \bar{K}_\alpha  \eta ) + i(\bar{\Omega
}^A \gamma _\mu  \bar{K}z \eta ) \nonumber
\\
 \delta \Omega ^A &=& - \frac{1}{2K\bar{z}} \left[ (\sigma A)^A + \Phi
_a{}^A \Phi _a{}^B (\sigma A)^B
\right] \eta  - i \hat{D} \Phi _a{}^A \bar{\tau }^a \eta  \label{s1} \\
 \delta \lambda _\alpha  &=& \frac{1}{4\bar{K}z} \varepsilon _{\alpha
\beta } z^\beta  (\sigma A) \Phi _a \bar{\tau }^a \eta  - i
\hat{D} z^\alpha  \eta  \nonumber \\
 \delta \Phi _a &=& (\bar{\Omega } \bar{\tau }^a \eta ) \qquad \delta
\bar{z}_\alpha  = 2(\bar{\lambda }_\alpha  \eta )
 \nonumber
\end{eqnarray}

   Now we can switch on a non-abelian gauge interactions. As usual, we
will assume that fields $\Phi _a{}^A$, $\Omega ^A$ ¨ $A_\mu {}^A$ are
transformed
under the adjoint representation of some gauge group $G$ with
structure constants $f^{ABC}$. In this, we have to replace all the
derivatives by covariant ones, e.g.:
\begin{eqnarray}
 \partial _\mu  \Phi _a{}^A &\to & {\cal{D}}_\mu  \Phi _a{}^A = \partial
_\mu  \Phi _a{}^A - f^{ABC} \Phi _a{}^B
A_\mu {}^C \nonumber  \\
 \partial _\mu  \Omega ^A &\to & {\cal{D}}_\mu  \Omega ^A = \partial _\mu
\Omega ^A - f^{ABC} \Omega ^B A_\mu {}^C  \\
 A_{\mu \nu }{}^A &=& \partial _\mu  A_\nu {}^A - \partial _\nu  A_\mu
{}^A - f^{ABC} A_\mu {}^B A_\nu {}^C
\nonumber
\end{eqnarray}
Then, in order to restore the invariance of the Lagrangian spoiled by
such replacement, one have to introduce the following additional terms
to the Lagrangian:
\begin{eqnarray}
 \Delta L &=& f^{ABC} \{ \frac{1}{24} \bar{\Psi }_\mu  \sigma ^{\mu \nu }
\bar{\Gamma }^{ABC}
(z\bar{K}) \Psi _\nu  + \frac{i}{12} (\bar{\Psi } \gamma ) \Gamma ^{ABC}
K^\alpha  \lambda _\alpha  +
\frac{1}{2} \bar{\lambda }_\alpha  K^\alpha  \Sigma ^{AB} \Omega ^C +
\nonumber \\
 && + \frac{i}{4} (\bar{\Psi } \gamma ) \Sigma ^{AB} (\bar{z}K) \Omega ^C
- \frac{1}{2}
\bar{\Omega }^A (\bar{z}K) \bar{\tau }^a \Phi _a{}^B \Omega ^C +
\frac{1}{24} \bar{\Omega }^D
\bar{\Gamma }^{ABC} (\bar{z}K) \Omega ^D \} - \nonumber   \\
 && - |\bar{z}K|^2 \left\{ \frac{1}{4} (f^{ABC} \Phi _a{}^A \Phi _b{}^B)^2
+ \frac{1}{6} (f \Phi _a \Phi _b \Phi _c)^2 \right\} ,
\end{eqnarray}
as well as complete the supertransformation laws with:
\begin{eqnarray}
 \delta '\Psi _\mu  &=& - \frac{i}{12} \gamma _\mu  f^{ABC} \bar{\Gamma
}^{ABC} (z\bar{K}) \eta
\nonumber \\
 \delta '\Omega ^A &=& - \frac{1}{2} (z\bar{K}) \{ f^{ABC} + \Phi _a{}^A
\Phi _a{}^D
f^{DBC} \} \Sigma ^{BC} \eta   \\
 \delta '\lambda _\alpha  &=& - \frac{1}{12} \varepsilon _{\alpha \beta }
z^{\beta } (\bar{z}K) f^{ABC}
\bar{\Gamma }^{ABC} \eta   \nonumber
\end{eqnarray}
Here we introduce the notations:
\begin{equation}
 \Sigma ^{AB} = \Phi _a{}^A \Phi _b{}^B \Sigma ^{ab} \qquad \Gamma ^{ABC}
= \Phi _a{}^A \Phi _b{}^B
\Phi _c{}^C \Gamma ^{abc}
\end{equation}
where matrices $(\Gamma ^{abc})_{ij}$, which are antisymmetric on $a,b,c$
and symmetric on $i,j$, are determined by the relation:
\begin{eqnarray}
 (\Gamma ^{[abc]})_{(ij)} &=& (\tau ^{[a})_{ik} (\bar{\tau }^b)^{kl} (\tau
^{c]})_{lj}
\label{m1}
\end{eqnarray}

   As in the $N=3$ supergravity case, global symmetry group $O(6,m)$
is non-compact, so one can consider a lot of different gaugings in this
theory. The first class consists of usual non-abelian gaugings with
the gauge groups like $O(3)\otimes O(3)\otimes H$, $O(3)\otimes
O(2,1)\otimes H$, $O(3)\otimes O(3,1)\otimes H$
and so on, where $H$ is some compact group. In general such a gauging
leads to the non-zero value of a cosmological term, but in some cases
cosmological term can be fine tuned to zero by ajusting the values of
different gauge coupling constant. One of the most interesting models
of this type is the one constructed in \cite{Wag88}, which gives $N=4
\to  N=1$ breaking.

   The second class of models corresponds to the possibility to have a
gauge group which contains translations as well as rotations. An
example of such gauging was constructed some time ago \cite{Por89} and
appeared to be the first (as far as we know) model giving spontaneous
supersymmetry breaking with two arbitrary scales. But the gravitini
masses turn out to be $\mu _1 = \mu _2 = (m_1+m_2)/2$ and $\mu _3 = \mu _4
=
(m_1-m_2)/2$, so that only $N=4 \to  N=2$ partial Higgs effect is
possible.

   The last possibility is the "abelian" gaugings corresponding to
pure translations similar to the model we have constructed for the
$N=3$ supergravity \cite{Zin91}. But in the $N=4$ case such models
give non-zero value of a cosmological term.

\section{Dual version}

   In this section we are going to construct the dual version which
will be the generalization of the hidden sector \cite{Zin87} to the
case of arbitrary number of vector multiplets. The crucial point is
that as the scalar field geometry remains unchanged we may use all
the terms without vector fields from the previous section. Let us
start by rewriting the corresponding parts of the Lagrangian
(\ref{l1}) and supertransformations (\ref{s1}) in terms of
independent spinor and scalar fields. For that purpose we introduce a
kind of light cone variables:
\begin{equation}
 \Phi _a{}^A = (x_a{}^m + E_{am}, x_a{}^m - E_{am}, \phi _a{}^{\hat{A}} ),
\qquad a,m = 1,2,...6
\end{equation}
Now, by introducing a new field
\begin{equation}
\pi ^{mn} = (E^{-1})^{am} x_a{}^n - (E^{-1})^{an} x_a{}^m
\end{equation}
we can solve the constraint (\ref{c2}) for $x_a{}^m$:
\begin{equation}
 x_a{}^m = \frac{1}{4} (2\delta _a{}^m + \Phi _{a\hat{A}}
\Phi _b{}^{\hat{A}}) (E^{-1})^{bm} + \frac{1}{2} E_{an} \pi ^{nm},
\end{equation}
Besides, in order all the scalar fields kinetic terms to be diagonal,
one have to make a change $\Phi _a{}^{\hat{A}} \to  E_{am} \Phi
^{m\hat{A}}$.

  Analogously, we solve the constraints (\ref{c2}) for spinor fields
introducing new fields $\Omega ^A = (\xi ^m + \chi _m, \xi ^m - \chi _m,
\Omega ^{\hat{A}})$.
This gives:
\begin{equation}
\xi ^m = -(E^{-1})^{am} x_a{}^n \chi _n + \frac{1}{2} \Phi ^{m\hat{A}}
\Omega _{\hat{A}}.
\end{equation}
Here two changes of variables $\chi _m \to  \frac{1}{\sqrt{2}} E_{am}\chi
^a$ and
$\Omega  \to  (\Omega  + \Phi ^m \chi _m)$ are necessary to have canonical
kinetic terms for
spinors.

   Now we are ready to rewrite all the terms without vector fields in
our new variables. Let us start with the hidden sector, the part of
the Lagrangian without matter fields $\Phi ^{m\hat{A}}$ and $\Omega
^{\hat{A}}$:
\begin{eqnarray}
L_1 &=& - \frac{1}{2} R + \frac{1}{2} D_\mu  z D_\mu  \bar{z} +
\frac{1}{2}
(S^+_\mu )^2 + \frac{1}{2} (P_\mu )^2 + \nonumber \\
&& + \frac{i}{2} \varepsilon ^{\mu \nu \alpha \beta }\bar{\Psi }_\mu
\gamma _5 \gamma _\nu  \left\{ D_\alpha  + \frac{1}{4}
(S_{\alpha }^- + P_\alpha )_{ab} \Sigma ^{ab} \right\} \Psi _\beta  +
\nonumber \\
&& + \frac{i}{2} \bar{\chi }^a \gamma ^\mu  \left\{ \delta _{ab} D_\mu  -
(S_\mu ^- -P_\mu )_{ab}
+ \frac{1}{4} \delta _{ab} (S_{\mu }^- + P_\mu )_{cd} \Sigma ^{cd}
\right\} \chi ^b +
\nonumber \\
&& + \frac{i}{2} \bar{\lambda }_\alpha  \gamma ^\mu  \left\{ D_\mu  -
\frac{1}{4} (S_\mu ^- +
P_\mu )_{ab} \Sigma ^{ab} \right\} \lambda _\alpha  - \nonumber \\
&& - \frac{1}{2} \bar{\lambda }_\alpha  \gamma ^\mu  \gamma ^\nu  D_\nu
z^\alpha  \Psi _\mu  - \frac{1}{2} \bar{\chi }^a
\gamma ^\mu  \gamma ^\nu  (S^+_\nu  + P_\nu )_{ab} \bar{\tau }^b \Psi _\mu
 \label{l2}
\end{eqnarray}
where we have introduced the following notations:
\begin{eqnarray}
(S^{\pm }_\mu )_{ab} &=& \frac{1}{2} (\partial _\mu  E_{am} (E^{-1})_b{}^m
\pm  (a\leftrightarrow b))
\nonumber  \\
(P_\mu )_{ab} &=& E_{am} \partial _\mu  \pi ^{mn} E_{bn}
\end{eqnarray}
The corresponding part of the supertransformations (\ref{s1}) has the
form:
\begin{eqnarray}
\delta \Psi _\mu  &=& 2D_\mu  \eta  + \frac{1}{2} (S^-_\mu  + P_\mu )_{ab}
\Sigma ^{ab} \eta  \nonumber \\
\delta \chi ^a &=& -i \gamma ^\mu  (S^+_\mu  + P_\mu )_{ab} \bar{\tau }^b
\eta  \qquad \delta \lambda _\alpha  = -i \gamma ^\mu
D_\mu  z^\alpha  \eta  \nonumber \\
\delta \bar{z}_\alpha  &=& 2(\bar{\lambda }_\alpha  \eta ) \qquad \delta
E_{am} = (\bar{\chi }^b E_{bm}
\bar{\tau }^a \eta ) \label{s2} \\
\delta \pi ^{mn} &=& \frac{1}{2} (\bar{\chi }^a \left[ (E^{-1})_a{}^m
(E^{-1})_b{}^n - (E^{-1})_b{}^m (E^{-1})_a{}^n \right] \bar{\tau }^b \eta
)
\nonumber
\end{eqnarray}
Note, that all the derivatives are the $U(1)$-covariant ones and
contain $(\bar{z} \partial _\mu  z)$ terms in according to (\ref{d1}).

    Part of the Lagrangian (\ref{l1}) with matter fields looks like:
\begin{eqnarray}
L_2 &=& \frac{1}{2} E_{am} E_{an} \partial _\mu  \Phi ^m \partial _\mu
\Phi ^n + \frac{i}{2}
\bar{\Omega } \gamma ^\mu  \left\{ D_\mu  + \frac{1}{4} (S_\mu ^- + P_\mu
)_{ab} \Sigma ^{ab}
\right\} \Omega  -  \nonumber \\
&& - i \bar{\chi }^a \gamma ^\mu  E_{am} \partial _\mu  \Phi ^m \Omega  -
\frac{1}{2} \bar{\Omega } \gamma ^\mu  \gamma ^\nu
E_{am} \partial _\nu  \Phi ^m \bar{\tau }^a \Psi _\mu    \label{l3}
\end{eqnarray}
and the corresponding supertransformations:
\begin{eqnarray}
\delta \Omega  &=& - i \gamma _\mu  E_{am} \partial _\mu  \Phi ^m\
bar{\tau }^a \eta  \qquad
\delta \Phi ^m = \bar{\Omega } (E^{-1})_{am} \bar{\tau }^a \eta
\label{s3}  \\
\delta \pi ^{mn} &=& - \frac{1}{2} \bar{\Omega } [\Phi ^m (E^{-1})_a{}^n -
\Phi ^n
(E^{-1})_a{}^m] \bar{\tau }^a \eta  \nonumber
\end{eqnarray}
In this, the expression for the $(P_\mu )_{ab}$ takes the form:
\begin{equation}
(P_\mu )_{ab} = E_{am} [\partial _\mu  \pi ^{mn} + \frac{1}{2} (\Phi ^m
\stackrel{\leftrightarrow } \partial _\mu
\Phi ^n)] E_{bn}
\end{equation}

   The essential part of the dual version is the hidden sector
\cite{Zin87}. Note that our current notations differ from whose used in
\cite{Zin87}. The main difference is that for the non-linear $\sigma
$-model
$SU(1,1)/U(1)$ we use doublet of scalar fields $z_\alpha $, satisfying
the constraint (\ref{c1}). Besides, let us denote twelve vector fields
as $(A_\mu )_m{}^{\hat{\alpha }}$, $\hat{\alpha } = 1,2$. In this, part of
the
Lagrangian containing vector fields will have the form:
\begin{eqnarray}
L_3 &=& - \frac{1}{4} U^{a}_{\mu \nu }U^{*a}_{\mu \nu } - \frac{1}{2} \pi
^{mn}
(\tilde{A}_{\mu \nu })_m{}^{\hat{\alpha }} \varepsilon _{\hat{\alpha
}\hat{\beta }} (A_{\mu \nu })_n{}^{\hat{\beta }}
- \frac{i}{4\sqrt{2}} \bar{\chi }^a \gamma ^\mu  (\sigma U)^a \Psi _\mu  -
\nonumber   \\
&& - \frac{1}{4\sqrt{2}} \bar{\Psi }_\mu  (U_{\mu \nu }^a - \gamma _{5}
\tilde{U}_{\mu \nu }^a)
\bar{\tau }^a \Psi _\nu  - \frac{i}{8\sqrt{2}} \bar{\lambda }_\alpha
\gamma ^\mu  \varepsilon _{\alpha \beta } z^\beta  (\sigma U^*)^a
\bar{\tau }^a \Psi _\mu  - \nonumber \\
&& - \bar{\lambda }_\alpha  \varepsilon ^{\alpha \beta } \bar{z}_\beta
(\sigma U)^a \chi ^a + \frac{1}{8\sqrt{2}}
\bar{\chi }^a (\sigma U^*)^b \bar{\tau }^b \chi ^a \label{l4}
\end{eqnarray}
and the appropriate part of the supertransformations:
\begin{eqnarray}
\delta \Psi _\mu  &=& \frac{i}{4\sqrt{2}} (\sigma U^*)^a \tau ^a \gamma
_\mu  \eta   \nonumber  \\
\delta (A_\mu )_m{}^{\hat{\alpha }} &=& \frac{E_{ma}}{\sqrt{2}} \left\{
(\bar{\Psi }_\mu
\bar{\tau }^a z^\alpha  G_\alpha {}^{\hat{\alpha }} \eta  ) + i (\bar{\chi
}^a \gamma _\mu  z^\alpha
G_\alpha {}^{\hat{\alpha }} \eta  ) - i (\bar{\lambda }_\alpha  \gamma
_\mu  \bar{\tau }^a G_\alpha {}^{\hat{\alpha }} \eta  )
\right\} \nonumber  \\
\delta \chi ^a &=& \frac{1}{2\sqrt{2}} (\sigma U)^a \eta  \qquad \delta
\lambda _\alpha  =
\frac{1}{4\sqrt{2}} \varepsilon _{\alpha \beta } z^\beta  (\sigma U^*)^a
\bar{\tau }^a \eta  \label{s4}
\end{eqnarray}
Here we introduced the following notation:
\begin{equation}
U_{\mu \nu }^a = (E^{-1})^{ma} z^\alpha  \bar{H}_\alpha {}^{\hat{\alpha }}
(A_{\mu \nu })_m{}^{\hat{a}},
\end{equation}
while matrices $G$ and $H$ have the form:
\begin{equation}
G_\alpha {}^{\hat{\alpha }} = \frac{1}{\sqrt{2}} \pmatrix{ 1 & \gamma _5
\cr1 & - \gamma _5
\cr},  \qquad  H^{\alpha \hat{\alpha }} = \frac{1}{\sqrt{2}} \pmatrix { 1
& - \gamma _5
\cr 1 & \gamma _5 \cr},
\end{equation}
They satisfy the following useful relations:
\begin{eqnarray}
 \varepsilon _{\alpha \beta } \bar{G}^{\beta \hat{\alpha }} = G_{\alpha
}{}^{\hat{\alpha }},  \qquad  \varepsilon ^{\alpha \beta }
\bar{H}_\beta {}^{\hat{\alpha }} &=& - H^{\alpha \hat{\alpha }}, \qquad
\bar{H}_\alpha {}^{\hat{\alpha }}
G_\beta {}^{\hat{\alpha }} = - \varepsilon ^{\hat{\alpha }\hat{\beta }}
\nonumber  \\
 \bar{G}^{\alpha \hat{\alpha }} G_\alpha {}^{\hat{\beta }} = - \gamma _5
\varepsilon ^{\hat{\alpha }\hat{\beta }},
\qquad  H^{\alpha \hat{\alpha }} \bar{H}_\alpha {}^{\hat{\beta }} &=& -
\gamma _5 \varepsilon ^{\hat{\alpha }\hat{\beta }},
\qquad H^{\alpha \hat{\alpha }} G_\alpha {}^{\hat{\beta }} = \delta
^{\hat{\alpha }\hat{\beta }}
\end{eqnarray}

   Now let us turn to the matter fields. Having in our disposal all
the terms without vector fields it is a relatively easy task to
complete the Lagrangian using ordinary Noether procedure. It turns out
that the only additional terms to the Lagrangian are:
\begin{eqnarray}
 L_4 &=& \frac{1}{2} \bar{\Omega }^{\hat{A}} \frac{1}{|K\bar{z}|^2}
\{ (\sigma V)^{\hat{A}} + (\sigma C)^{\hat{A}} \} K^\alpha  \lambda
_\alpha  + \frac{1}{8\sqrt{2}}
\bar{\Omega }^{\hat{A}} (\sigma \bar{U})^a \bar{\tau }^a \Omega ^{\hat{A}}
+ \nonumber  \\
&& + \frac{i}{4} \bar{\Omega }^{\hat{A}} \gamma ^\mu  \frac{1}{K\bar{z}}
\{
(\sigma V)^{\hat{A}} + (\sigma C)^{\hat{A}} \} \Psi _\mu  - \nonumber \\
&& - \frac{1}{4|K\bar{z}|^2} (V_{\mu \nu }^{\hat{A}} +
C_{\mu \nu }^{\hat{A}})^2 + \frac{\gamma _5}{4} \left[
\frac{\bar{M}z}{\bar{K}z}
- h.c. \right] (C_{\mu \nu }^{\hat{A}})^2 -  \nonumber  \\
&& - \frac{1}{4} (V_{\mu \nu }^{\hat{A}} + 2 C_{\mu \nu }^{\hat{A}})
\tilde{W}_{\mu \nu }^{\hat{A}} \label{l5}
\end{eqnarray}
and to the supertransformations, correspondingly:
\begin{eqnarray}
 \delta \Omega ^{\hat{A}} &=& - \frac{1}{2K\bar{z}} \{ (\sigma
C)^{\hat{A}} +
(\sigma V)^{\hat{A}} \} \eta  \nonumber \\
 \delta C_{\mu }^{\hat{A}} &=& i (\bar{\Omega }^{\hat{A}} \gamma _\mu
(\bar{K}z) \eta ) + i
(\bar{\chi }^a \gamma _\mu  E_{am} \Phi ^{m\hat{A}} (\bar{K}z) \eta ) +
\label{s5}  \\
 && + (\bar{\Psi }_\mu  \bar{\tau }^a E_{am} \Phi ^{m\hat{A}} (\bar{K}z)
\eta ) - i
(\bar{\lambda }_\alpha  \gamma _\mu  \bar{K}_\alpha  \bar{\tau }^a E_{am}
\Phi ^{m\hat{A}} \eta ) ,  \nonumber
\end{eqnarray}
where $K_{\alpha }$ ¨ $M_{\alpha }$ are the same as in the (\ref{l1}) and
we have
introduced:
\begin{eqnarray}
V_{\mu \nu }^{\hat{A}} &=& \frac{1}{\sqrt{2}} \{ (K\bar{z}) U_{\mu \nu }^a
+
(\bar{K}z) U_{\mu \nu }^{*a} \} E_{am} \Phi ^{m\hat{A}} \nonumber  \\
W_{\mu \nu }^{\hat{A}} &=& \frac{\gamma _5}{\sqrt{2}} \{ (\bar{K}z) U_{\mu
\nu }^{*a} -
(K\bar{z}) U_{\mu \nu }^a \} E_{am} \Phi ^{m\hat{A}}
\end{eqnarray}

\section{Supersymmetry breaking}

   Thus we have managed to construct the dual version of the $N=4$
supergravity --- vector multiplets system which indeed generalize our
hidden sector \cite{Zin87} to the case of arbitrary number of vector
multiplets. As we have already mentioned, the scalar field geometry
are the same as in the ordinary version, but the global symmetry of
the Lagrangian and the transformation properties of the vector fields
differ in these two models. Now, instead of $O(6,m)$ group, we have
$O(m-6)\otimes GL(6)$ global symmetry, as well as fifteen translations
$\pi ^{mn} \to  \pi ^{mn} + \Lambda ^{mn}$, under which all the vector
fields are
inert. Therefore, by analogy with the $N=3$ case, one can try to
introduce the masses of the vector fields $(A_\mu )_m{}^{\hat{\alpha }}$
by
using $\pi ^{mn}$ as Goldstone fields and replacing
\begin{equation}
 \partial _\mu  \pi ^{mn} \to  \partial _\mu  \pi ^{mn} -
M^{mnp}_{\hat{\alpha }} (A_\mu )_p{}^{\hat{\alpha }}
\label{d2}
\end{equation}
The only difficulty that arises here is related to
the "axion" term $\frac{1}{2} \pi ^{mn} \varepsilon _{\hat{\alpha
}\hat{\beta }}
(\tilde{A}_{\mu \nu })_m{}^{\hat{\alpha }} (A_{\mu \nu })_n{}^{\hat{\beta
}}$ in the
Lagrangian. It can be completed to a gauge invariant expression only
provided the tensors $M^{mnp}_{\hat{\alpha }}$ are fully skew-symmetric in
their indices. Besides, as will become clear below, one has to impose
the condition
\begin{equation}
M_{\hat{\alpha }}^{mnp} = - \frac{1}{6} \varepsilon ^{\hat{\alpha
}\hat{\beta }} \varepsilon ^{mnpqrs}
M_{\hat{\beta }}^{qrs}  \label{c3}
\end{equation}
In this the gauge invariant combination has the form
\begin{equation}
 \frac{1}{2} \pi ^{mn} \varepsilon _{\hat{\alpha }\hat{\beta }}
(\tilde{A}_{\mu \nu })_m{}^{\hat{\alpha }}
(A_{\mu \nu })_n{}^{\hat{\beta }} + \frac{2}{3} M^{mnp}_{\hat{\alpha }}
\varepsilon _{\hat{\beta }\hat{\gamma }} (A_\mu )_p{}^{\hat{\alpha }}
(A_\nu )_m{}^{\hat{\beta }}
(\tilde{A}_{\mu \nu })_n{}^{\hat{\gamma }}
\end{equation}
which corresponds to the self-interaction of \it abelian \rm massive
vector fields!

As usual, substitution (\ref{d2}) breaks the invariance of the
Lagrangian under the supertransformations. To compensate for
this noninvariance, it is necessary to complete the Lagrangian
of the hidden sector with
\begin{eqnarray}
L' &=& - \frac{1}{6} (\bar{z} \bar{K})^{\hat{\alpha }}
Q_{abc}^{\hat{\alpha }}
Q_{abc}^{\hat{\beta }} (zK)^{\hat{\beta }} + \frac{1}{72} \varepsilon
^{abcdef}
\varepsilon _{\hat{\alpha }\hat{\beta }} Q_{abc}^{\hat{\alpha }}
Q_{def}^{\hat{\beta }} + \nonumber \\
 && + \frac{1}{24\sqrt{2}} \bar{\Psi }_\mu  \sigma ^{\mu \nu }
(zK)^{\hat{\alpha }}
Q_{abc}^{\hat{\alpha }} \bar{\Gamma }^{abc} \Psi _\nu  +
\frac{i}{4\sqrt{2}} \bar{\Psi }_\mu
\gamma ^\mu  (\bar{z} \bar{})^{\hat{\alpha }} Q_{abc}^{\hat{\alpha }}
\Sigma ^{ab} \chi ^c +
\nonumber \\
 && + \frac{i}{12\sqrt{2}} \bar{\Psi }_\mu  \gamma ^\mu
Q_{abc}^{\hat{\alpha }}\Gamma ^{abc}
\bar{K}^{\alpha \hat{\alpha }} \lambda _\alpha  - \frac{1}{2\sqrt{2}}
\bar{\chi }^a
Q_{abc}^{\hat{\alpha }} (\bar{z}\bar{K})^{\hat{\alpha }} \bar{\tau }^b
\chi ^c + \nonumber
\\
 && + \frac{1}{24\sqrt{2}} \bar{\chi }^d Q_{abc}^{\hat{\alpha }}
\bar{\Gamma }^{abc}
(\bar{z} \bar{K})^{\hat{\alpha }} \chi ^d + \frac{1}{2\sqrt{2}}
\bar{\lambda }_\alpha
\bar{K}^{\alpha \hat{\alpha }} Q_{abc}^{\hat{\alpha }} \Sigma ^{ab} \chi
^c \label{l6}
\end{eqnarray}
where the following notation is introduced:
\begin{equation}
Q_{abc}^{\hat{\alpha }} = E_{am} E_{bn} E_{cp} M^{mnp}_{\hat{\alpha }}
\end{equation}
In this the total Lagrangian is invariant under supertransformations
with additional terms of the form
\begin{eqnarray}
\delta '\Psi _\mu  &=& - \frac{i}{12\sqrt{2}} \gamma _\mu
(zK)^{\hat{\alpha }} Q_{abc}
\bar{\Gamma }^{abc} \eta  \nonumber \\
\delta '\chi ^a &=& - \frac{1}{2\sqrt{2}} (zK)^{\hat{\alpha }}
Q^{*abc}\Sigma ^{bc}\eta    \\
\delta '\lambda _\alpha  &=& - \frac{1}{12\sqrt{2}} \varepsilon _{\alpha
\beta }z^\beta  (\bar{z}\bar{K})^{\hat{\alpha }}
Q_{abc}^{\hat{\alpha }} \bar{\Gamma }^{abc} \eta   \nonumber
\end{eqnarray}

   As for the matter fields, it turns out that no new terms in the
supertransformations appear, while the Lagrangian have to be completed
with the only new term
\begin{equation}
\frac{1}{24\sqrt{2}} \bar{\Omega }^{\hat{A}} (\bar{z}\bar{K})^{\hat{\alpha
}}
Q_{abc}^{\hat{\alpha }} \bar{\Gamma }^{abc} \Omega ^{\hat{A}}
\end{equation}

So far, we have not given the tensors $M^{mnp}_{\hat{\alpha }}$
explicitly. Clearly, in the general case one gets a family of
tensors related via the O(6) - transformations. It would be
convenient to choose a parameterization in which the gravitino mass
matrix is diagonal. Then in the concrete representation of
$\tau $-matrices (see the Appendix) one has
\begin{equation}
M^{mnp}_1 = m_1 \delta _1^{[m} \delta _2^n \delta _3^{p]} + m_2 \delta
_1^{[m} \delta _5^n \delta _6^{p]}
+ m_3 \delta _2^{[m} \delta _4^n \delta _6^{p]} - m_4 \delta _3^{[m}
\delta _4^n \delta _5^{p]}
\end{equation}
Here the gravitino masses turn out proportional to the combinations
\begin{equation}
 \left( \begin{array}{c} m_1 + m_2 + m_3 + m_4 \\ m_1 - m_2 - m_3 +
m_4 \\ m_1 - m_2 + m_3 - m_4 \\ m_1 + m_2 - m_3 - m_4 \end{array}
\right)
\end{equation}
Analysing potential (the first two terms in the (\ref{l6})), one can
show that its minimum corresponds to $<z_0>=1$, $<z_1>=0$,
$(E_{am}=\delta _{am})$. Besides, by virtue of the properties (\ref{c3})
and (\ref{c5}), the value of the potential at the minimum equals zero
which corresponds to the absence of a cosmological term.

Hence, the theory constructed really admits a spontaneous
supersymmetry breaking without a cosmological term and with \it four
\rm arbitrary scales, including the partial super-Higgs effect $N=4
\to  N=3$, $N=4 \to  N=2$ and $N=4 \to  N=1$. But, due to the constraint
(\ref{c3}), which is essential for the absence of the cosmological
term, vacuum expectation value of $(\bar{z} \bar{K})^{\hat{\alpha }}
Q_{abc}^{\hat{\alpha }}$ turns out to be zero and the spontaneous
supersymmetry breaking does not lead to the appearance of the matter
field mass terms.

\section{Conclusion}

   Thus we have seen that in the "minimal" version for $N=4$
supergravity with matter till now no one have managed to construct a
fully satisfactory model i.e. the model which gives spontaneous
supersymmetry breaking with different scales, a zero value for
cosmological term without fine tuning and the possibility to have
partial super-Higgs effect $N=4 \to  N=1$. The dual version, constructed
in this paper, indeed has all these desired properties but failed to
generate soft breaking terms for matter fields. Let us stress,
however, that in both cases there exist a lot of possible gaugings,
which have not been fully explored so far.

\vskip 1cm

{\Large \bf Acknowledgments}

   Work supported by International Science Foundation grant RMP000 and
by Russian Foundation for Fundamental Researches grant 94-02-03552.

\newpage
\appendix
\section{}
We have used the following $\tau $-matrix representation
\begin{eqnarray}
\tau ^1 &=& \left( \begin{array}{cccc} 0&0&0&1 \\ 0&0&1&0 \\ 0&-1&0&0 \\
-1&0&0&0 \end{array} \right) \qquad \tau ^4 = \left( \begin{array}{cccc}
0&0&0& \gamma _5 \\ 0&0& -\gamma _5& 0 \\ 0& \gamma _5 &0&0 \\ -\gamma _5
&0&0&0 \end{array}
\right) \nonumber \\
\tau ^2 &=& \left( \begin{array}{cccc} 0&0&-1&0 \\ 0&0&0&1 \\ 1&0&0&0 \\
0&-1& 0&0 \end{array} \right) \qquad  \tau ^5 = \left( \begin{array}{cccc}
0&0&\gamma _5&0 \\ 0&0&0&\gamma _5 \\ -\gamma _5&0&0&0 \\ 0&-\gamma _5&0&0
\end{array} \right)
\\
\tau ^3 &=& \left( \begin{array}{cccc} 0&1&0&0 \\ -1&0&0&0 \\ 0&0&0&1 \\
0&0&-1&0 \end{array} \right) \qquad \tau ^6 = \left( \begin{array}{cccc}
0&-\gamma _5&0&0 \\ \gamma _5&0&0&0 \\ 0&0&0&\gamma _5 \\ 0&0&-\gamma _5&0
\end{array} \right)
\nonumber
\end{eqnarray}
Here the matrices $\Gamma ^{abc}$ satisfy the relation
\begin{equation}
\Gamma ^{abc} = \frac{1}{6} \gamma _5 \varepsilon ^{abcdef} \Gamma ^{def}
] \label{c5}
\end{equation}
In this representation, the following matrices are diagonal:
\begin{eqnarray}
\Gamma ^{123} &=& \left( \begin{array}{cccc} 1&0&0&0 \\ 0&1&0&0 \\ 0&0&1&0
\\ 0&0&0&1 \end{array} \right) \qquad \Gamma ^{156} = \left(
\begin{array}{cccc} 1&0&0&0 \\ 0&-1&0&0 \\ 0&0&-1&0 \\ 0&0&0&1
\end{array} \right) \nonumber \\
\Gamma ^{246} &=& \left( \begin{array}{cccc} 1&0&0&0 \\ 0&-1&0&0 \\
0&0&1&0
\\ 0&0&0&-1 \end{array} \right) \qquad \Gamma ^{345} = \left(
\begin{array}{cccc} -1&0&0&0 \\ 0&-1&0&0 \\ 0&0&1&0 \\ 0&0&0&1
\end{array} \right)
\end{eqnarray}
as well as the matrices $\Gamma ^{456}$, $\Gamma ^{234}$, $\Gamma ^{135}$
and $\Gamma ^{126}$
related with them through relation (\ref{c5}).

\end{document}